\documentclass[sigconf]{acmart}
\AtBeginDocument{%
  }

\setcopyright{acmlicensed}
\copyrightyear{2026}
\acmYear{2026}
\acmDOI{XXXXXXX.XXXXXXX}

\acmConference[AIware '26]{the 3rd ACM International Conference on AI-Powered Software}{July 06--07, 2026}{Montreal, QC, Canada}
\acmISBN{978-1-4503-XXXX-X/26/06}

\usepackage{booktabs}
\usepackage{graphicx}
\usepackage{balance}
\usepackage{tabularx}
\usepackage{ragged2e}
\usepackage{placeins}
\usepackage{float}
\usepackage{xcolor}
\usepackage[most]{tcolorbox}
\newtcolorbox{rqsummary}[1]{
  enhanced jigsaw,
  colback=white,
  colframe=black,
  boxrule=0.6pt,
  width=\linewidth,
  sharp corners,
  top=1pt,
  bottom=1pt,
  left=5pt,
  right=5pt,
  before upper={\textbf{\small #1}\\[1pt]},
}
\newcommand{\CompactListSpacing}{%
  \setlength{\itemsep}{1pt plus 0.5pt minus 0.5pt}%
  \setlength{\parsep}{0pt}%
  \setlength{\topsep}{4pt plus 1pt minus 1pt}%
  \setlength{\partopsep}{0pt}%
}

\usepackage{microtype}

\begin{document}
\hypersetup{colorlinks=true,linkcolor=black,citecolor=black,urlcolor=ACMDarkBlue,filecolor=ACMDarkBlue}
\renewcommand\UrlFont{\color{ACMDarkBlue}\rmfamily}
\makeatletter
\def\orcid#1{\unskip\ignorespaces%
  \protected\def\orcidsite{https://orcid.org/}%
  \IfBeginWith{#1}{http}{%
    \expandafter\gdef\csname typeset@author\the\num@authors\endcsname##1{%
      \href{#1}{\textcolor{ACMDarkBlue}{##1}}}}{%
    \expandafter\gdef\csname typeset@author\the\num@authors\endcsname##1{%
      \href{\orcidsite#1}{\textcolor{ACMDarkBlue}{##1}}}}}
\makeatother

\title{Collaborator or Assistant? How AI Coding Agents Partition Work Across Pull Request Lifecycles}
\author{Young Jo(seph) Chung}
\orcid{0009-0002-9656-2632}
\affiliation{%
  \institution{University of Toronto}
  \department{Faculty of Information}
  \city{Toronto}
  \state{Ontario}
  \country{Canada}
}
\email{jo.chung@utoronto.ca}

\author{Safwat Hassan}
\orcid{0000-0001-7090-0475}
\affiliation{%
  \institution{University of Toronto}
  \department{Faculty of Information}
  \city{Toronto}
  \state{Ontario}
  \country{Canada}
}
\email{safwat.hassan@utoronto.ca}
\renewcommand{\shortauthors}{Chung and Hassan}
\begin{abstract}
When AI coding agents open branches and submit pull requests (PRs),
two questions co-determine oversight design: who \emph{starts} the
work (operational agency) and who \emph{authorizes} its completion
(merge governance).
We characterize tools along a \emph{Collaborator--Assistant
spectrum} in how they redistribute initiative, oversight, and
endorsement, while merge governance remains predominantly human
across five tools (OpenAI, Copilot, Devin, Cursor, Claude Code).
We analyze 29,585~PR lifecycles using an Initiator~$\times$~Approver
taxonomy with six interaction scenarios; lifecycle reconstruction
supplies the \emph{how} behind those roles.
Collaborator tools (Cursor, Devin, Copilot) concentrate operational
initiative in agents that open and carry PR work forward, with humans
retaining review and endorsement on the path to merge; Assistant tools
(OpenAI, Claude) leave task direction primarily with humans and supply
bounded support within human-led workflows.
Across the spectrum, agency and governance \emph{decouple}:
Collaborator workflows are $\geq$96\% agent-initiated, yet terminal
merge authority remains almost exclusively human, with
agent-classified approvers confined to a small fraction of PRs.
Where automation executes a merge, logs record the executor but not
the decision-maker, marking a boundary of observation.
We contribute the taxonomy, per\nobreakdash-tool state machines, and a
replication package for research on automation, oversight, and
governance in PR workflows.
\end{abstract}

\ccsdesc[500]{Software and its engineering~Collaboration in software development}
\ccsdesc[300]{Software and its engineering~Software development process management}

\keywords{AI coding agents, human-AI collaboration, merge governance, software engineering, pull request workflows, empirical study}

\maketitle

\section{Introduction}

When AI coding agents submit pull requests, two coupled questions
co-determine oversight design: do these agents function as
\emph{collaborators} or as \emph{assistants} on operational work,
and who holds \emph{merge governance} when they do?
The first concerns \emph{operational agency}---who initiates and
carries the PR forward; the second concerns \emph{governance
authority}---who may authorize its completion.
These are not the same dimension: agents now create branches, write
code, run tests, and submit PRs
autonomously~\cite{li2025aiteammates}, yet a tool can concentrate
initiation on agents while reserving merges for humans, or the
reverse.
We use \emph{Collaborator} and \emph{Assistant} as behavioral
workflow paradigms, not merely as labels for the actor who submits the
first commit.
In the Collaborator paradigm, operational initiative is redistributed
toward the AI system: the agent proposes or opens the work, structures
the PR artifact, and carries the lifecycle far enough that humans
primarily exercise oversight, review, and endorsement.
In the Assistant paradigm, the human remains the practical director of
the work: the agent supplies bounded capability, such as generation,
revision, or recommendation, within a workflow whose goals and
trajectory remain human-led.
We distinguish \emph{endorsement} from \emph{accountability} along this
axis.
Endorsement refers to clear acceptance of a specific PR outcome, such
as review, approval, or merge authorization; accountability refers to
who is responsible for the merged code after it enters the codebase.
The human team may still be accountable for that code even when no
person gives a clear approval and the final step is handled by
repository rules such as branch protection or auto-merge.
This distinction draws on agency and delegation research that treats
initiative, autonomy, and responsibility as separable but
accountability-relevant dimensions of human--AI work
arrangements~\cite{baird2021,raisch2021automation,fugen2021}.
Together they shape tool design, team
workflows~\cite{raisch2021automation},
and trust calibration~\cite{okamura2020trust}.
Empirically, they also vary by tool: in some ecosystems, nearly every
agent-generated PR passes through human review; in others, most
PRs reach the main branch without a recorded review step.
Without lifecycle data that separates initiation from terminal
authorization, review allocation, trust calibration,
and tool selection rely on anecdote rather than
evidence.

Recent studies report what AI-agent PRs achieve but not how
they get there.
Gao et al.\ find reduced review engagement in bot-associated
PRs~\cite{gao2026autopilot};
Rahman et al.\ evaluate task-level acceptance
rates~\cite{rahman2026tasklevel};
Pinna et al.\ compare merge rates stratified by task
type~\cite{pinna2026comparing};
and Agarwal et al.\ contrast IDE assistants with autonomous
agents on productivity metrics~\cite{agarwal2026ai}.
These studies measure \emph{what happened} (merge rates, comment
counts) rather than \emph{who did what, when}: none reconstructs
the full PR lifecycle or separates the actor who initiates work
from the actor who authorizes its completion.

More broadly, empirical research on human--AI collaboration in
software engineering has examined discrete
interactions (code suggestions, commit patterns, or review
comments~\cite{treude2025developers}) and predominantly studied
single phases rather than full
lifecycles~\cite{treude2012empirical}.
Related evidence from outside repository mining also shows that
large-scale user feedback on Gen-AI software surfaces recurring
concerns (e.g., performance, output quality, and policy/censorship),
but this lens captures user-perceived app behavior rather than PR
lifecycle governance~\cite{almulla2025understandingthechallenges}.
Without a lifecycle view, teams lack evidence-based guidance for
allocating oversight in AI-assisted development.

To address this gap, we draw on Sheridan and Verplank's levels of
automation~\cite{sheridan1978human} and process mining
analysis~\cite{vanderaalst2016process} to study 29,585~PR
lifecycles across five AI coding
tools~\cite{li2025aidev}.
\textbf{Main finding.}
Tools fall along a \textbf{Collaborator--Assistant spectrum} on
initiation and review routing (strong tool--scenario association:
Cram\'er's V~$= 0.50$), while merge governance stays concentrated in
human hands.
Our Initiator~$\times$~Approver taxonomy separates \emph{operational
agency} (who starts work) from \emph{governance authority} (who
merges); these dimensions decouple empirically: Collaborator tools
show $\geq$96\% agent-initiated PRs, yet $<$0.1\% receive
agent-authorized merges.
Agents do much of the operational work but hold almost none of the
merge authority, a ``junior teammate'' pattern that directly informs
review allocation (Section~\ref{sec:discussion}).
The dataset concentrates temporally (94\% of PRs from
2025~Q2--Q3), making this a cross-sectional snapshot of an
emerging practice rather than a longitudinal trend analysis.
In the rare cases where agents \emph{do} hold merge authority,
event logs record who executed the merge but not who made the
governance decision, exposing a measurement boundary at the limits
of what lifecycle mining can reveal.

\textbf{Research Questions.}

\textbf{RQ1:} How do humans and AI agents partition work across
complete PR lifecycles?

\textbf{RQ2:} How do collaboration workflows and phase transitions
differ across AI coding tools?

\textbf{RQ3:} What do automation-authorized merges reveal about
the boundaries of the Initiator~$\times$~Approver taxonomy?

We address these in Section~\ref{sec:results}.

\textbf{Contributions.} We make three contributions:
\begin{itemize}\CompactListSpacing
    \item \textbf{Analytical Framework.} A lifecycle model (three
    phases, two terminal outcomes) paired with an
    Initiator~$\times$~Approver taxonomy of six interaction scenarios
    (Section~\ref{sec:methodology}).
    \item \textbf{Empirical Characterization.} Analysis of 29,585~PRs
    with five per-tool state machines quantifying workflow dynamics,
    plus path analysis of all automation-authorized merges
    (Section~\ref{sec:results}).
    \item \textbf{Replication Package.} The complete dataset,
    analysis pipeline, and generated artifacts.\footnote{\url{https://doi.org/10.6084/m9.figshare.31343038}}
\end{itemize}

\section{Background and Related Work}
\label{sec:background}

\subsection{AI Coding Agents and Levels of Automation}

AI-assisted programming evolved from syntax autocomplete to
context-aware code generation (e.g., GitHub Copilot), and the
current generation (Devin, Cursor, Claude Code) marks a further
shift toward \emph{agentic} behavior: autonomous task execution
rather than suggestion~\cite{li2025aiteammates}.
Sheridan and Verplank's levels of automation~\cite{sheridan1978human}
frame this trajectory: predictive assistants operate at
Levels~2--3 (computer suggests, human chooses), whereas agentic
systems advance toward Level~5 (execute after human approval) or
Level~6+ (execute autonomously).
Li et al.\ characterize this transition as ``AI teammates'' in
SE~3.0, where agents participate as dialectical partners rather
than passive tools~\cite{li2025aiteammates}.

\textbf{Difference from outcome-centric evaluations.}
Much prior work on AI-assisted PRs is outcome-centric, comparing
tools by merge rate, time-to-merge, or comment
volume~\cite{li2025aidev,pinna2026comparing,rahman2026tasklevel}.
These metrics are informative but ambiguous about \emph{who did
what} across the lifecycle.
Our contribution is process-centric: we reconstruct PR
trajectories over phases (Created, Review, Revision, terminal
outcome) and classify each PR into one of six
Initiator~$\times$~Approver scenarios (S1--S6).
Researchers can then surface distinctions that aggregated outcome
metrics conflate; for example, high acceptance may reflect
agent-initiated work accepted by others~(S1) versus
human-initiated work selectively submitted and later
self-merged~(S4).

\textbf{Difference from design-centric tool taxonomies.}
Another strand classifies tools by intended interaction mode
(e.g., IDE assistants versus autonomous
agents)~\cite{li2025aiteammates,agarwal2026ai}.
Such capability-based taxonomies describe affordances but do not
necessarily reflect observed usage, especially when tools support
multiple modes.
Our Collaborator--Assistant spectrum is a behavioral
classification grounded in event traces (initiation and terminal
authorization), disentangling operational agency (who initiates
work) from governance authority (who merges).

\textbf{Complement to findings on reduced oversight.}
Reports of ``silent merges'' in AI-associated PRs highlight
reduced oversight, but outcome- and comment-level analyses
were not designed to localize where oversight is
bypassed~\cite{gao2026autopilot,rahman2026tasklevel}.
By modeling lifecycle transitions, we provide a structural
mechanism: in the Assistant paradigm, PRs frequently resolve
directly from initial development to merge without recorded entry
into a review phase.
This complements social and behavioral interpretations by showing
that the difference is also a workflow-level pattern observable in
event sequences.

\subsection{Pull-based development and GitHub evidence}

Foundational empirical work established pull-based development as a
systematic research topic on GitHub-scale data~\cite{gousios2014pull}
and cataloged methodological pitfalls of mining GitHub~\cite{kalliamvakou2014github}.
We follow the first line by modeling lifecycle-level PR dynamics with
explicit roles; we follow the second by treating RQ3 as a concrete case
where event logs underdetermine governance (\emph{who decided} versus
\emph{who executed} the merge).

\subsection{Human-AI Collaboration and Trust}

The \emph{automation-augmentation paradox}~\cite{raisch2021automation}
frames a central tension: AI can replace human tasks (automation)
or enhance human capabilities (augmentation).
\emph{Distributed cognition} theory~\cite{hollan2000distributed}
posits that cognitive work distributes across actors and artifacts,
but distributed \emph{activity} does not imply distributed
\emph{responsibility}, a distinction central to PR workflows
where agents initiate work yet humans retain merge authority.

This tension makes trust calibration critical: overtrust produces
automation complacency, while undertrust limits productivity
gains~\cite{okamura2020trust,glikson2020trust}.
Roychoudhury et al.\ propose ``programming with trust,'' earned
through transparency and verification~\cite{roychoudhury2025trust}.
Evans and Stanovich~\cite{evans2013dualprocess} and Halpern~\cite{halpern2014thought}
appear in HCI-related work as sources of language about defaults,
deliberate override, and monitoring in review-like settings; we cite
them only as \emph{optional vocabulary} for readers who want a
familiar parallel to the observed split between PRs merged with
little review and PRs routed through explicit review.
The empirical claims in this paper do not depend on that parallel
(or on any cognitive model of developers).

\subsection{Process Mining in Software Engineering}

Process mining treats event logs as traces through state
machines~\cite{vanderaalst2016process}, with applications spanning
software configuration
management~\cite{rubin2014process,rubin2015system} and
CI/CD workflows~\cite{nogueira2021cicd}.
Classical process discovery induces models from data; we instead
define the lifecycle model \emph{a priori} from domain knowledge
and adopt the process mining \emph{analysis} framework, computing
transition probabilities and median sojourn times to quantify
human-agent workflow dynamics.

\section{Methodology}
\label{sec:methodology}

\subsection{Overview}

\begin{figure}[t]
\centering
\includegraphics[width=\linewidth]{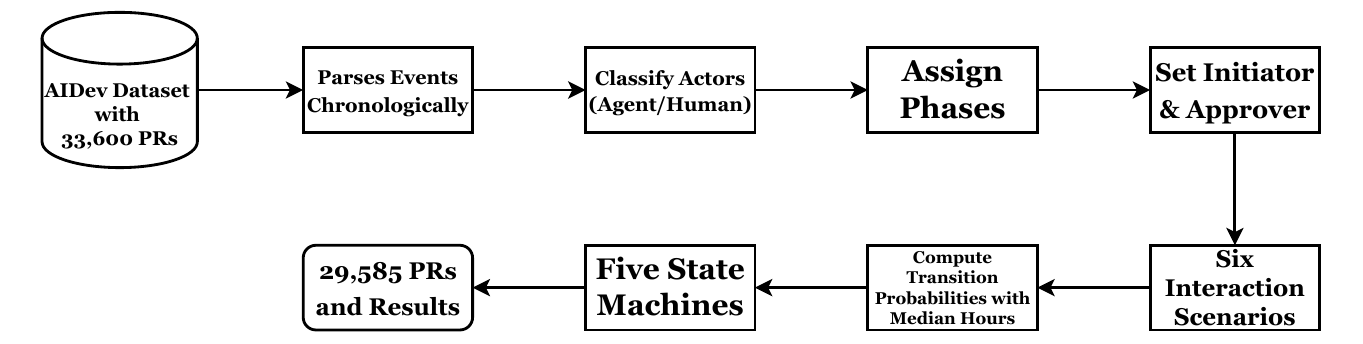}
\caption{High-level overview of the data pipeline.}
\Description{Flowchart: AIDev dataset input; parse events; classify actors; assign phases; set initiator and approver; classify S1--S6 scenarios; compute transition probabilities; generate state machines; produce results.}
\label{fig:pipeline-overview}
\end{figure}

We analyze \textbf{29,585~PR workflows with terminal outcomes}
from the AIDev dataset across five AI coding tools
(Figure~\ref{fig:pipeline-overview}).
After excluding incomplete timelines
(Section~\ref{sec:exclusion}), we treat each PR as a complete
lifecycle and compute phase transition probabilities and median
hours by tool, yielding five state machines
(Section~\ref{sec:results}).

\textbf{Pipeline.} For each PR the pipeline:
\begin{enumerate}\CompactListSpacing
    \item Parses events chronologically.
    \item Classifies actors as Agent or Human via a two-step
    heuristic.
    \begin{enumerate}\CompactListSpacing
    \item If GitHub's \texttt{actor.type} field equals
    \texttt{"Bot"}, the actor is classified as Agent.
    \item For all remaining actors the lowercased login is
    checked against the patterns: \texttt{bot},
    \texttt{copilot}, \texttt{devin}, \texttt{cursor},
    \texttt{codex}, \texttt{openai}, \texttt{claude}.
    A login that contains any listed substring is classified as
    \textbf{Agent}; otherwise the actor is classified as
    \textbf{Human}.
    \end{enumerate}
    The pattern check covers missing or non-standard
    \texttt{actor.type} values.
    Validation against events with available \texttt{actor.type}
    shows that residual errors are rare (0.36\%) and conservative for
    our main contrast: they classify some humans as Agent, but do not
    classify true Bots as Human.
    We return to the remaining small-sample risk in
    Section~\ref{sec:threats}.
    \item Assigns phases (PR created~$\to$
    Review~$\rightleftharpoons$ Revision~$\to$ terminal).
    \item Sets initiator (first \texttt{committed} actor) and
    approver (merge actor).
    \item Classifies the PR into one of six scenarios (S1--S6;
    Table~\ref{tab:taxonomy}).
\end{enumerate}
Aggregation steps:
\begin{enumerate}\CompactListSpacing
    \setcounter{enumi}{5}
    \item \textbf{Transition probabilities.} Computes P(To$|$From) and
    median hours per phase and transition by tool.
    \item \textbf{Generate state machines.} Generates workflow diagrams
    per tool with phase nodes and transition edges.
    \item \textbf{Results.} Produces six-scenario distribution,
    paradigm classification, and state machine figures.
\end{enumerate}
All steps are deterministic and reproducible.

\subsection{Dataset}

We use the AIDev dataset with 33,600~PRs from GitHub repositories with
AI coding agent activity~\cite{li2025aidev}.
The dataset includes a curated subset of \textbf{qualification PRs}
(repositories with $>$100 GitHub stars) with enriched event data
(review comments, commit-level diffs, and full PR timelines;
Table~\ref{tab:dataset}).
We abbreviate the OpenAI tool family (GPT-4o, o1, etc.)\ as
``OpenAI'' and Claude Code as ``Claude.''
Throughout, ``Copilot'' refers to GitHub Copilot's agent and
Workspace variants captured in the AIDev dataset, distinct
from Copilot's inline-suggestion IDE mode, which is not
represented in this corpus and would likely exhibit
Assistant-like initiation behavior.

Repository selection in the AIDev dataset is
\emph{tool-driven}: repositories enter the dataset when at
least one PR involves a recognized AI coding agent actor
(e.g., \texttt{copilot[bot]}, \texttt{devin-ai[bot]}).
This strategy captures repositories \emph{where tools are
used}, not a random sample of all GitHub repositories; it follows
a tool-driven sampling strategy used in empirical SE studies of
emerging GitHub phenomena~\cite{li2025aidev,kalliamvakou2014github},
but one that limits generalizability to the population of
AI-tool-adopting repositories.

Each timeline contains \texttt{committed},
\texttt{review\_requested}, \texttt{reviewed},
\texttt{commented}, and \texttt{merged} or \texttt{closed} events.
Not all PRs have complete lifecycle data; we apply exclusion
criteria to obtain an analytic sample (Section~\ref{sec:exclusion},
Table~\ref{tab:dataset}).

\subsubsection{Exclusion Criteria}
\label{sec:exclusion}

We exclude PRs that cannot be classified into our workflow
taxonomy. Two criteria apply: (1)~no \texttt{committed} event
(15~PRs; initiator undefined) and
(2)~no \texttt{merged} or \texttt{closed} event (4,000~PRs;
11.9\%; no terminal outcome).
Table~\ref{tab:dataset} shows the composition by tool before
and after applying these criteria.
The analytic sample comprises \textbf{29,585~PRs} with terminal
outcomes. All statistics and figures use this sample.

\begin{table}[t]
\caption{Dataset composition by tool. \textbf{Total}: raw count.
\textbf{No-Commit}: excluded (no \texttt{committed} event).
\textbf{Incomplete}: excluded (no \texttt{merged} or \texttt{closed} event).
\textbf{Included}: analytic sample with terminal outcomes.}
\label{tab:dataset}
\small
\begin{tabular}{lrrrr}
\toprule
\textbf{Tool} & \textbf{Total} & \textbf{No-Commit} & \textbf{Incomplete} & \textbf{Included} \\
\midrule
Cursor & 1,541 & 1 & 241 & 1,299 \\
Devin & 4,829 & 5 & 595 & 4,229 \\
Copilot & 4,971 & 2 & 2,107 & 2,862 \\
Claude Code & 459 & 1 & 120 & 338 \\
OpenAI & 21,800 & 6 & 937 & 20,857 \\
\midrule
\textbf{TOTAL} & \textbf{33,600} & \textbf{15} & \textbf{4,000} & \textbf{29,585} \\
\bottomrule
\end{tabular}
\end{table}

\subsection{Workflow Lifecycle Model}

\begin{figure}[t]
\centering
\includegraphics[width=\linewidth]{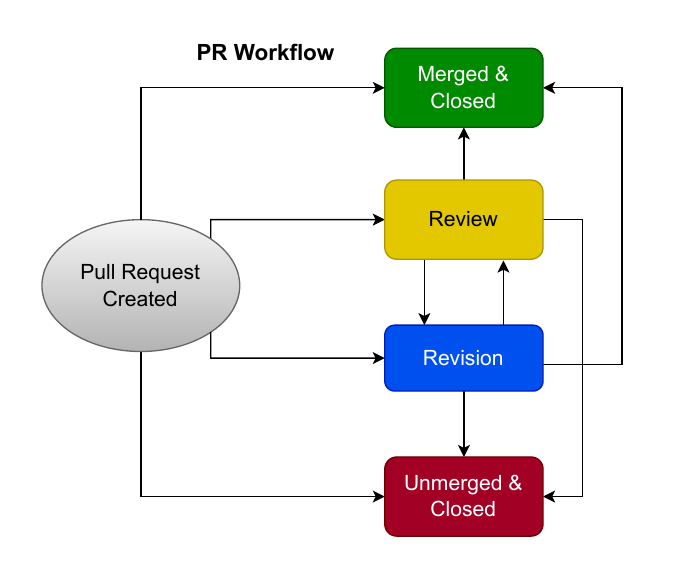}
\caption{PR workflow lifecycle: phases and terminal outcomes.}
\Description{State machine diagram showing transitions between Pull Request Created, Review, and Revision phases towards Merged and Closed or Unmerged and Closed terminal outcomes.}
\label{fig:pr-workflow}
\end{figure}

We define a PR lifecycle with three non-terminal phases and two
terminal outcomes (Figure~\ref{fig:pr-workflow},
Table~\ref{tab:phase-def}).

We scan each timeline chronologically and map events to phases.
Some PRs close and reopen (e.g., 46~Devin, 27~OpenAI); the final
terminal event is definitive.
State machines exclude intermediate Closed transitions.
Revision cycle counts (REVIEW~$\rightleftharpoons$~REVISION) are
computed separately and available in the replication package.

\begin{table}[t]
\caption{Phase definitions for the PR lifecycle model.
\textsuperscript{*}``PR created'' denotes the \emph{phase}
(the interval from PR opening to first review); the instantaneous
event of opening a PR is referred to as ``PR creation.''
State machine diagrams label this phase ``Pull Request Created.''}
\label{tab:phase-def}
\centering
\small
\begin{tabularx}{\linewidth}{@{}l>{\raggedright\arraybackslash}X@{}}
\toprule
\textbf{Phase / Outcome} & \textbf{Definition} \\
\midrule
PR created\textsuperscript{*} & Entry: PR timeline opens at the first recorded event. The PR remains here while work is being prepared and before recorded review begins. Exit: first \texttt{review\_requested} or \texttt{reviewed} after $\geq$1 \texttt{committed}. \\
Review & Entry: first review-routing event after a commit, either \texttt{review\_requested} or \texttt{reviewed}. The PR remains here while approval or rejection is pending. Exit: \texttt{CHANGES\_REQUESTED}, merge, or close. \\
Revision & Entry: reviewer submits a \texttt{reviewed} event with state \texttt{CHANGES\_REQUESTED}. The PR remains here while the author or agent responds through commits and renewed review requests. Exit: return to Review, merge, or close. \\
Merged \& Closed & Terminal outcome entered when the PR is merged. \\
Unmerged \& Closed & Terminal outcome entered when the PR is closed without merge. \\
\bottomrule
\end{tabularx}
\end{table}

\subsection{Interaction Scenario Taxonomy}

We operationalize the Collaborator--Assistant distinction through six
scenarios defined by initiator (first \texttt{committed} actor) and
approver (Table~\ref{tab:taxonomy}).
The taxonomy separates \textbf{operational agency} (who starts and
carries work into the PR lifecycle) from \textbf{governance authority}
(who authorizes completion through merge).

Merged PRs are split by approver (Human or Agent); non-merged PRs
collapse into Not-Merged.
The approver is a derived classification from the merge actor, not a
raw governance field in the dataset.
Approver classification uses the same bot-pattern heuristics as
initiator classification.
GitHub timelines often attribute merges to generic bots
(e.g., \texttt{github-actions[bot]}), so S2/S5 are
\emph{automation-authorized}, not agent-tool-authorized.

We classify tools as \emph{Collaborator} when S1+S2+S3 exceeds
90\%, or \emph{Assistant} when S4+S5+S6 exceeds 90\%.
This threshold is an empirical operationalization of the broader
workflow concepts introduced above.
Collaborator tools make the AI system the practical source of
initiative while humans retain review, endorsement, and merge
governance; Assistant tools keep the human as task director while the
agent contributes intermediate activities such as generation,
revision, or recommendation within human-led workflows.
The taxonomy therefore measures a sharper axis than initiation alone:
where operational agency sits, whether endorsement is visible or
thin, and how far accountability remains assigned to humans despite
automation in the lifecycle.

\begin{table}[t]
\caption{Six interaction scenarios (Initiator $\times$ Approver).}
\label{tab:taxonomy}
\centering
\small
\begin{tabularx}{\linewidth}{@{}cl>{\raggedright\arraybackslash}X@{}}
\toprule
\textbf{ID} & \textbf{Type} & \textbf{Criteria} \\
\midrule
& \textbf{Collaborator} & Agent-led operational agency; tools with $\geq$90\% of PRs in S1+S2+S3. \\
S1 & Agent-Init + Human-Approved & initiator=Agent, approver=Human \\
S2 & Agent-Init + Agent-Approved & initiator=Agent, approver=Agent \\
S3 & Agent-Init + Not-Merged & initiator=Agent, closed without merge \\
\midrule
& \textbf{Assistant} & Human-led task direction with agent support; tools with $\geq$90\% of PRs in S4+S5+S6. \\
S4 & Human-Init + Human-Approved & initiator=Human, approver=Human \\
S5 & Human-Init + Agent-Approved & initiator=Human, approver=Agent \\
S6 & Human-Init + Not-Merged & initiator=Human, closed without merge \\
\bottomrule
\end{tabularx}
\end{table}

Collaborator tools (S1--S3) combine agent-led generation and PR
lifecycle movement with human oversight at review and merge;
Assistant tools (S4--S6) reverse this, with humans initiating and
directing work and agents supporting within human-led workflows.
(Section~\ref{sec:background}).

\section{Results}
\label{sec:results}

\subsection{RQ1. How do humans and AI agents partition work across complete PR lifecycles?}

\textbf{Approach.} We classify each PR by initiator and approver
into six scenarios (S1--S6; Table~\ref{tab:taxonomy}), yielding a
contingency table of 5 tools $\times$ 6 scenarios.
We apply two tests.
First, a $\chi^2$ test of independence: under the null
hypothesis, tool and scenario are unrelated (each tool would
show the same scenario distribution).
We reject the null if observed counts deviate significantly
from this expectation.
Second, Cram\'er's V measures effect size: how strongly
knowing the tool predicts the scenario (0~= no association;
1~= perfect).
For a 5~$\times$~6 table (df~$= 4$), Cohen's ``large'' threshold
is $V \geq 0.25$~\cite{cohen1988power}.

\begin{figure*}[t]
\centering
\includegraphics[width=0.9\linewidth]{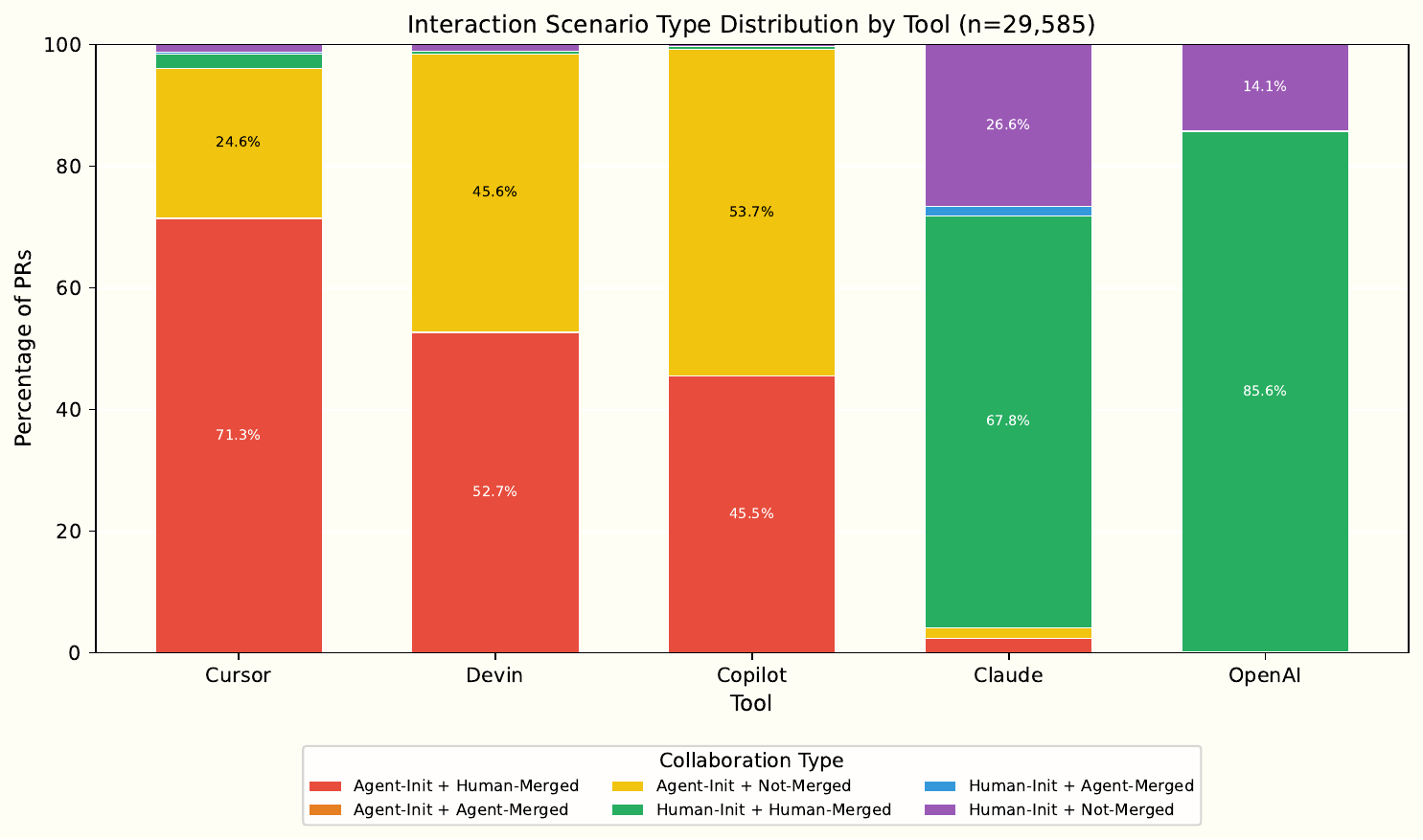}
\caption{Interaction scenario distribution by tool
(29,585 included PRs).
Stacked bars show percentage per tool in S1--S6.
Copilot has 0.8\% Human-Init (S4+S6; 23~PRs).}
\Description{Stacked bar chart showing the percentage of PRs in each of the six interaction scenarios (S1-S6) for each tool, highlighting the Collaborator vs Assistant paradigms.}
\label{fig:collab-dist}
\end{figure*}

\textbf{Results.} Figure~\ref{fig:collab-dist} shows the
scenario distribution.
The $\chi^2$ test rejects independence
($\chi^2 = 29{,}817$, df~$= 20$, $p < 0.001$).
Cram\'er's V~$= 0.50$ (twice the large-effect threshold for
df~$= 4$) indicates tool identity strongly predicts which
scenario pattern a PR follows.

Tools cluster along a spectrum.
Collaborator tools (Cursor, Devin, Copilot) show $\geq$96\%
agent-initiated PRs (S1+S2+S3).
Assistant tools (OpenAI, Claude) show $\geq$95.6\%
human-initiated PRs (S4+S5+S6).
OpenAI dominates the Assistant cluster
($n > 20{,}800$ vs.\ Claude $n = 338$); the pattern is best
supported as an OpenAI--Collaborator contrast.

Direct resolution rates vary within paradigms.
OpenAI resolves 76.5\% of PRs directly from initial development;
Claude resolves 37.6\%, closer to Cursor (34.6\%) than to OpenAI.
Claude's small sample ($n = 338$) yields wide confidence intervals
(e.g., S4: [45.4, 54.6]\%).
Claude's patterns are therefore presented as tentative; the main
contributions (spectrum, $V = 0.50$, state machines, governance
gap) hold across the four larger-sample tools.

\noindent\textbf{Operational agency vs.\ governance authority.}
Initiation and approval \emph{decouple}.
Agents start $\geq$96\% of Collaborator PRs, yet S2
(Agent-Init + Agent-Approved) accounts for $<$0.1\% across
all tools.
Agents initiate heavily but hold near-zero merge authority.

\begin{rqsummary}{Summary of RQ 1}
Tools fall into Collaborator ($\geq$96\% agent-initiated) and
Assistant ($\geq$95.6\% human-initiated) paradigms; tool identity
strongly predicts scenario pattern (Cram\'er's V~$= 0.50$).
S1 and S4 (human-approved) dominate merged outcomes; S2 and S5
(agent-approved) remain rare.
Governance stays human-centric regardless of who starts the work.
\end{rqsummary}

\noindent\textbf{Semi-autonomous agent pattern (S2).}
S2 totals 14~PRs ($\leq$0.1\% per tool).
The merge actor is typically a generic automation bot, not the
coding agent itself (Section~\ref{sec:methodology}).

\subsection{RQ2. How do collaboration workflows and phase transitions differ across AI coding tools?}

\begin{figure}[t]
\centering
\includegraphics[width=\linewidth]{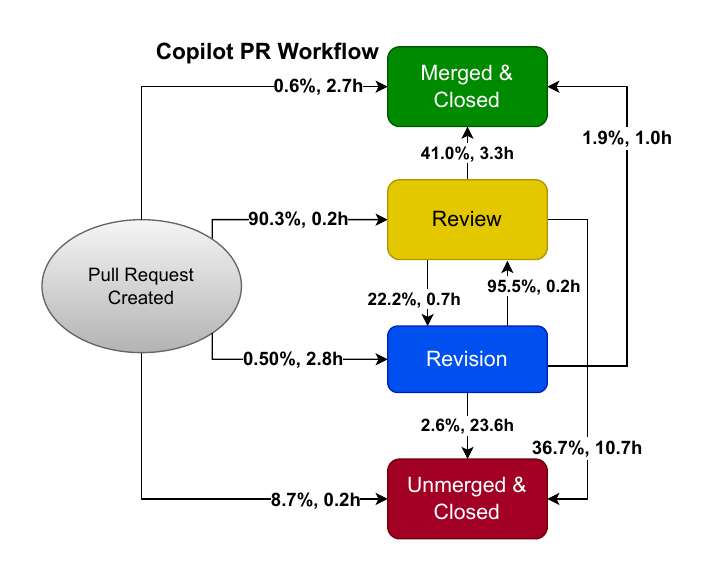}
\caption{Copilot PR workflow: phase transition probabilities and median hours per state.}
\Description{State machine diagram for Copilot showing phase transition probabilities and median hours spent in each state.}
\label{fig:sm-copilot}
\end{figure}

\begin{figure}[t]
\centering
\includegraphics[width=\linewidth]{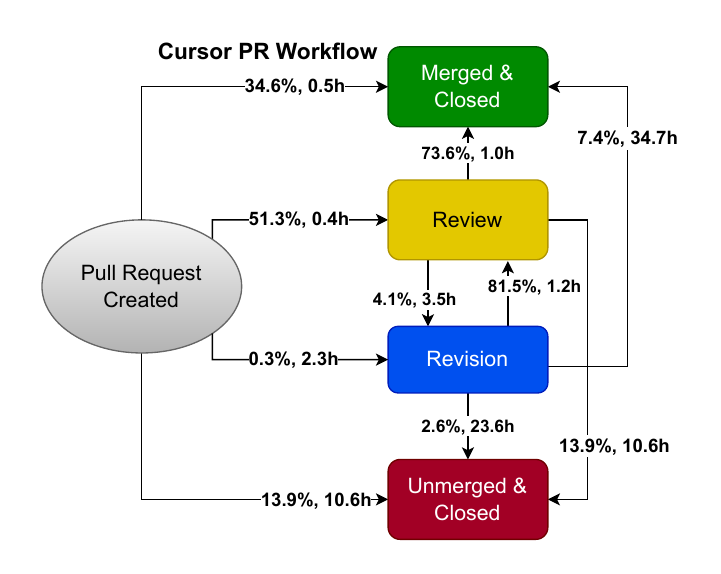}
\caption{Cursor PR workflow: phase transition probabilities and median hours per state.}
\Description{State machine diagram for Cursor showing phase transition probabilities and median hours spent in each state.}
\label{fig:sm-cursor}
\end{figure}

\begin{figure}[t]
\centering
\includegraphics[width=\linewidth]{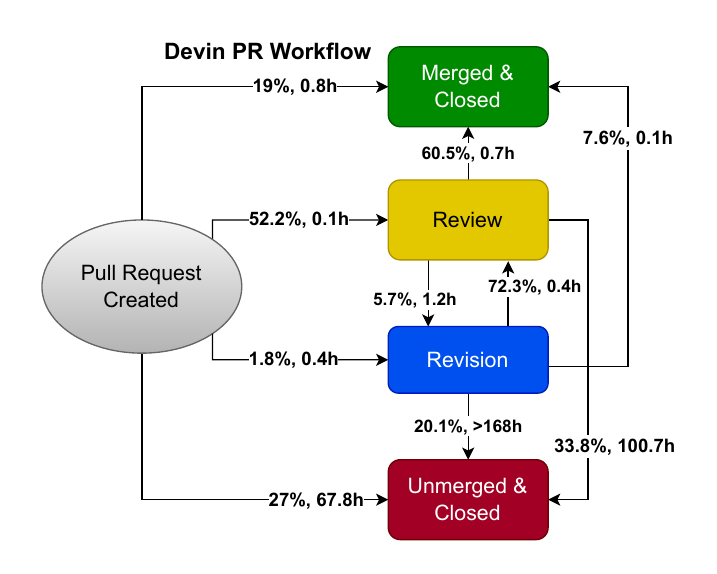}
\caption{Devin PR workflow: phase transition probabilities and median hours per state.}
\Description{State machine diagram for Devin showing phase transition probabilities and median hours spent in each state.}
\label{fig:sm-devin}
\end{figure}

\begin{figure}[t]
\centering
\includegraphics[width=\linewidth]{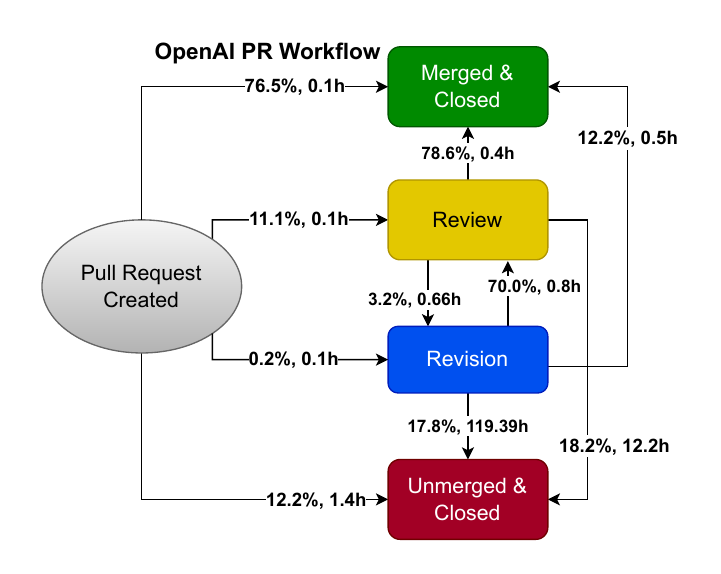}
\caption{OpenAI PR workflow: phase transition probabilities and median hours per state.}
\Description{State machine diagram for OpenAI showing phase transition probabilities and median hours spent in each state.}
\label{fig:sm-openai}
\end{figure}

\begin{figure}[t]
\centering
\includegraphics[width=\linewidth]{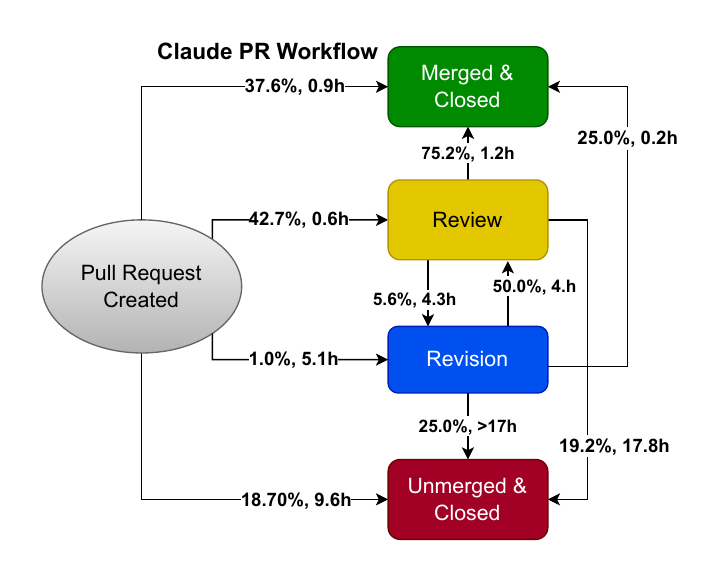}
\caption{Claude PR workflow: phase transition probabilities and median hours per state.}
\Description{State machine diagram for Claude showing phase transition probabilities and median hours spent in each state.}
\label{fig:sm-claude}
\end{figure}

\textbf{Approach.} We compute transition probabilities
$P(\text{To} \mid \text{From})$ and median hours per state by
tool, producing five state machines
(Figures~\ref{fig:sm-copilot}--\ref{fig:sm-claude}).
Nodes show median hours; edges show transition probability and
median hours.

\textbf{Results.} Collaborator tools route most PRs through
Review: Copilot~90.3\%, Cursor~51.3\%, Devin~52.2\%.
Assistant tools resolve directly: OpenAI~76.5\%, Claude~37.6\%.
Revision~$\to$~Review return rates are higher for Collaborators
(Copilot~95.5\%, Devin~72.3\%), indicating substantive revision
loops.
Median Review time: Copilot~3.0\,h, Devin~2.0\,h vs.\
OpenAI~0.7\,h.
Time to Unmerged~\&~Closed varies sharply: Devin~67.8\,h vs.\
Copilot~0.2\,h and OpenAI~1.4\,h.
High rejection latency in Devin indicates extended deliberation;
rapid closure in Copilot and OpenAI indicates automated triage.

\begin{rqsummary}{Summary of RQ 2}
Collaborators route PRs through Review (Copilot~90.3\%,
Cursor~51.3\%, Devin~52.2\%) with substantive revision loops;
Assistants resolve directly (OpenAI~76.5\%, Claude~37.6\%).
Median hours reveal the contrast: Collaborators 2--3\,h in
Review vs.\ OpenAI~0.7\,h; time to Unmerged varies sharply
(Devin~67.8\,h vs.\ Copilot~0.2\,h, OpenAI~1.4\,h).
Claude has a split profile (37.6\% direct, closer to Cursor
than OpenAI).
\end{rqsummary}

To control for initiation-driven routing differences, we restricted
the sample to human-initiated PRs and compared review-routing rates.
When initiation is held constant (human-initiated PRs only), the
pattern persists: Collaborator tools still route most through Review
(Copilot~87.8\%, Devin~61.0\%, Cursor~57.1\%) while OpenAI
resolves 76.6\% directly to merge.
Claude shows an intermediate profile (43.2\% to Review,
37.8\% to Merged), reinforcing its split profile.
Wilson score 95\% confidence intervals confirm non-overlapping
ranges between paradigms: Copilot~[75.8, 94.3]\%,
Devin~[49.9, 71.2]\%, Cursor~[44.1, 69.2]\% versus
OpenAI~[10.7, 11.6]\%.
Even at the lower bounds, Collaborator tools exceed OpenAI's
upper bound by 3.8--6.5$\times$.
Remaining circularity risks are discussed in
Section~\ref{sec:threats}.

\subsection{RQ3. What do automation-authorized merges reveal about the boundaries of the taxonomy?}

\textbf{Approach.} S2 and S5 are the only scenarios with
Agent-classified approvers: 54~PRs total (S2:~14; S5:~40),
0.24\% of merged PRs.
We perform exhaustive path analysis of all 54, classifying each
by phase sequence.

\textbf{Results.}
We partitioned the 54~PRs by path.
In 33 cases (61\%), a human completed review before a bot account
executed the merge.
There the derived approver classification records the merge
\emph{mechanism} (automation performed the click), not the governance
\emph{decision} (the human reviewer had already accepted the change).
These PRs function as S1 or S4: the governance decision was human,
and the bot merely executed the merge.

The remaining 21~PRs (39\%) merged directly with no recorded
review phase.
Many repository setups can yield the same GitHub event pattern for
those merges.
Human-configured auto-merge after green checks, one-off waivers of
branch protection, and CI/CD merge automation can each produce a
bot-attributed merge in the log.
Timeline data alone do not distinguish those benign explanations from a
merge driven by genuinely autonomous agent governance.
We therefore treat 21 of 29,585~PRs (0.07\%) only as an \emph{upper}
bound on autonomous authority; without branch-protection and
merge-policy metadata, that bound is plausibly high.

\noindent\textbf{Tool concentration.}
S5 concentrates in OpenAI repositories (32/40; 80\%),
consistent with CI/CD auto-merge infrastructure.
S2 distributes evenly (Devin~5, OpenAI~5, Copilot~2, Cursor~2).

\noindent\textbf{Implication.}
S2/S5 rarity has at least three separable sources:
\emph{technical capability} (GitHub Actions can execute merges; the
infrastructure exists), \emph{tool default exposure}
(most coding agents do not by default surface merge authority
to the agent), and \emph{organizational trust} (where exposure
exists, teams may still withhold the affordance).
Our event logs localize the joint outcome but cannot separate
these layers (Section~\ref{sec:discussion}).
Analysts using the taxonomy can identify these scenarios and
track governance delegation as it evolves.

\begin{rqsummary}{Summary of RQ 3}
Of 54~PRs with Agent-classified approvers (S2+S5, 0.24\% of merged),
61\% passed through human Review before a bot executed the merge;
``Agent-Approved'' thus reflects merge \emph{mechanism}, not governance
\emph{decision}.
The remaining 39\% merged directly; the same event pattern can
reflect human-configured automation (auto-merge, protection
waivers, CI/CD) or genuinely autonomous agent governance, and
timelines alone do not separate those cases.
The taxonomy captures who \emph{executes} the merge but not who
\emph{decides}; resolving this requires repository-level metadata
unavailable in the dataset.
\end{rqsummary}

\section{Discussion}
\label{sec:discussion}

\subsection{Two Paradigms of Human-Agent Collaboration}

The Collaborator--Assistant spectrum reflects two distinct modes
of human-agent integration.

\textbf{Collaborator Paradigm.}
Cursor, Devin, and Copilot show agent-initiated PRs routed
through review, aligning with Raisch and Krakowski's
``augmentation'' model~\cite{raisch2021automation}: agents
generate output autonomously, and humans validate through formal
review.
Review routing rates (Section~\ref{sec:results}) confirm that deliberate oversight is structurally embedded in these workflows.

\textbf{Assistant Paradigm.}
OpenAI and Claude invert this pattern: human-initiated PRs with
higher direct resolution.
Phase transition analysis reinforces the contrast: Collaborators
show higher review routing and substantive revision loops, whereas
Assistants resolve directly.

\textbf{Repository governance as confound.}
These patterns, however, do not necessarily reflect developer
choice alone.
Direct resolution can reflect repository governance as much as
behavioral preference.
GitHub's \emph{branch protection rules}~\cite{github2024branchprotection}
are repository-level policies that require pull request reviews,
mandate passing status checks, and restrict who can merge; they shape
whether a PR receives review before closure.
Protected main or release branches often require approving reviews,
while feature branches may permit direct merges, and administrators
can bypass restrictions unless explicitly included.
Thus the same observed lifecycle can reflect tool behavior,
repository policy, or their interaction.

Our checks separate these interpretations where the data allow.
Among human-initiated PRs only, Wilson score 95\% confidence intervals
for review routing remain non-overlapping between paradigms, with
Collaborator tools exceeding OpenAI's upper bound by
3.8--6.5$\times$ (Section~\ref{sec:results}).
The within-repository control (Section~\ref{sec:threats}) confirms
that the initiation gap persists (96.8~percentage points), but
review-routing contrasts partly reverse.
The safest interpretation is therefore asymmetric: the
Collaborator--Assistant distinction is strongest for redistributed
initiative, while review routing also reflects repository adoption and
governance context.
Automation complacency~\cite{okamura2020trust,glikson2020trust} is
consistent with the data but not confirmed.
RQ3 supports the same caution: 61\% of automation-authorized merges
involved human review, so ``Agent-Approved'' often reflects
infrastructure, not autonomous endorsement.
OpenAI's 76.5\% direct resolution is consistent with accepting agent
output without deliberate review; team-level review norms may matter
as much as tool design.

\textbf{The Collaborator--Assistant spectrum.}
Claude has a \emph{split profile}: it meets the Assistant
threshold by initiation pattern ($\geq$95.6\% human-initiated),
yet its review behavior resembles Collaborator tools:
37.6\% direct resolution, closer to Cursor (34.6\%) than to
OpenAI (76.5\%).
This underscores that initiation mode and evaluation intensity
are separable dimensions, and that the Collaborator--Assistant
classification captures \emph{dominant tendencies} along a
spectrum rather than rigid categories.
Claude's classification as Assistant rests on its initiation
pattern ($\geq$95.6\% human-initiated) and should be interpreted
with caution given $n = 338$; replication with a larger Claude
corpus is needed before drawing firm conclusions about its
position on the spectrum.

\subsection{The Governance Boundary}

The ``junior teammate'' pattern describes a \emph{current
empirical state}, not an architectural constraint.
GitHub Actions can already submit reviews, approve changes, and
execute merges, so the infrastructure exists; branch protection
rules, tool defaults, and organizational trust determine whether
such authority is exposed and used.
RQ3's 54 automation-authorized merges are among the earliest
observed instances of end-to-end lifecycle automation in production
repositories, and their rarity reflects where organizations
currently draw the line between human and automated merge
authority.
As tools advance from Level~5 (execute after approval) toward
Level~6+ (execute autonomously)~\cite{sheridan1978human}, the
question shifts from \emph{how is work partitioned?} to
\emph{how should authority be delegated?}
Mined event logs do not answer this question. That gap is
the limitation that RQ3's measurement boundary makes explicit.

\subsection{Practical Implications}

The Collaborator--Assistant spectrum informs three operational
decisions.

\textbf{Review capacity planning.}
Collaborator tools generate agent-initiated PRs that route through
human review at high rates (Copilot~90.3\%, Cursor~51.3\%,
Devin~52.2\%).
Teams deploying these tools can use sojourn times (e.g.,
Copilot median 3.0\,h in Review) and revision return rates
(e.g., $P(\text{Revision} \mid \text{Review}) = 0.222$ for
Copilot) as sprint-planning inputs for review workload.

\textbf{Risk-based review strategies.}
In Assistant tool workflows, direct resolution dominates
(OpenAI~76.5\%).
Rather than mandating review for all PRs, teams may consider
risk-based strategies, such as reviewing PRs that exceed complexity
thresholds (e.g., file count, lines changed) while allowing
low-risk PRs to resolve directly.

\textbf{Graduated trust models.}
The governance gap (0.07\% autonomous merges) marks where
organizations currently draw the line on automated merge authority.
Tool designers and platform administrators can use this baseline
to calibrate graduated trust: auto-merge for low-risk,
well-tested changes and human approval for high-risk
modifications, rather than blanket autonomy or blanket
restriction.

\subsection{Future Research}

Three directions follow from the measurement boundary and the
Collaborator--Assistant spectrum.

\textbf{From mined logs to live API data.}
Resolving RQ3's measurement boundary requires data not present
in mined event logs: branch protection rules, required-reviewer
policies, merge-queue configurations, and bot permission scopes,
available through live platform APIs.
The AIDev dataset lacks branch protection metadata, and GitHub's
API requires admin access to query protection rules
(\texttt{403} without \texttt{administration=read} scope),
making retrospective estimation infeasible at scale.
Our proxy reasoning (if branch protection were the sole driver
of review routing, all tools in protected repositories would
show similar rates) is suggestive but incomplete, as tool
adoption and governance structures may co-vary.
Acquiring live API data (protection rules, merge-queue
configurations) is the most critical methodological priority.

\textbf{Quality and outcome analysis.}
Correlating S1--S6 with commit complexity, test coverage, and
post-merge defect rates would test whether direct resolution
trades off quality.

\textbf{Review content analysis.}
Examining reviewer requests in Collaborator revision loops
would distinguish substantive quality concerns from
mandatory process steps that add delay without improving the change.
Mixed-methods work (interviews, surveys) would address
\emph{why} teams choose direct resolution over formal
review~\cite{okamura2020trust,glikson2020trust}.

\section{Threats to Validity}
\label{sec:threats}

\textbf{Internal validity.}
Bot identification uses naming-pattern heuristics (0.36\% error
rate vs.\ \texttt{actor.type} metadata).
Timestamps reflect GitHub server time, not offline activity.
The principal threat is repository governance confounding (Section~\ref{sec:discussion}): direct resolution rates may partly reflect repository permission structures.

\textbf{External validity.}
The dataset spans 2018--2025 but concentrates temporally:
Q2~2025 accounts for 55.4\% of PRs, Q3 for 36.5\%, Q1 for
4.9\%, and Q4~2024 for 0.2\%.
This 94\% concentration in Q2--Q3~2025 reflects tools reaching
production maturity in early 2025; findings represent a
cross-sectional snapshot, not longitudinal trends.
Tool-specific repository selection introduces potential bias.
Copilot's exclusion rate (2,107/4,971; 42.4\%) warrants
scrutiny.
Median creation dates for included vs.\ excluded PRs differ by one day
(2025-06-25 vs.\ 2025-06-24), mitigating temporal bias; however,
63.3\% of excluded PRs (1,334/2,107) hit the 30-event API pagination cap,
truncating terminal events.
Exclusion reflects a \emph{data collection artifact}: excluded
PRs are more active PRs whose longer timelines exceeded the
pagination window.
The included sample may underrepresent complex, multi-event
lifecycles; this is a conservative bias, since those PRs likely had
richer review-revision cycling.
OpenAI comprises 70.5\% of the analytic sample.
To verify the spectrum is not a sample-size artifact, we
downsampled OpenAI to $n = 4{,}885$ across 10~random seeds: Cram\'er's~V remained
stable at 0.510~$\pm$~0.001 (range 0.5096--0.5108), and the
Collaborator--Assistant bifurcation held in all seeds.
As a within-repository control, we restricted analysis to 34~repositories where each Collaborator and Assistant tool had at least five PRs ($n = 2{,}197$ PRs).
Agent-initiated shares remained bifurcated (Collaborator~97.1\%, Assistant~0.3\%), a 96.8~percentage-point gap matching the full dataset---consistent with a tool-design interpretation of initiation.
The share of merges with a human approver reversed relative to the cross-repository pattern (Assistant~59.6\% vs.\ Collaborator~49.7\%), suggesting that RQ2 review-routing contrasts partly reflect repository adoption and governance, not only tool-intrinsic routing.

\textbf{Construct validity.}
The taxonomy captures \emph{who}, not \emph{why}.
A circularity risk exists: tools clustered by initiation rate are
then analyzed for review frequency, a potential downstream
consequence.
To mitigate this, we compared human-initiated PRs exclusively (Section~\ref{sec:results}): Collaborators still route most through review while OpenAI resolves directly, with non-overlapping Wilson 95\% confidence intervals.
Within-repository analysis partially addresses repository-level confounding for RQ2; initiation contrasts remain stable across specifications.
Revision counts do not capture revision quality.
RQ3 shows the Approver dimension conflates merge
\emph{mechanism} with governance \emph{decision}; resolving this
requires governance metadata unavailable in the dataset.

\textbf{Endorsement measurement.}
Our event-log measure of \emph{endorsement} records the
occurrence of a \texttt{reviewed} event, not its semantic depth;
a 30-second Looks Good To Me (LGTM) and a thorough code review are observationally
identical in the timeline data.
Distinguishing substantive review from cursory approval requires
content-level analysis (comment text, lines reviewed, and
change-request substance) that we leave to future work.
The endorsement axis introduced in Section~1 is therefore a
theoretical distinction that present logs can \emph{localize} but
not \emph{measure}.

\textbf{Content-level controls.}
We do not control for PR size, complexity, or task type.
If tool affordances systematically interact with PR
characteristics---for example, Collaborator tools generating
larger or more structurally complex changes---the review-routing
contrast in RQ2 could partly reflect content rather than tool
design.
Pinna et al.'s~\cite{pinna2026comparing} task-type stratification
represents the natural next step for disentangling these effects.

\enlargethispage{6\baselineskip}
\section{Conclusion}
\label{sec:conclusion}

29,585~PR lifecycles across five tools reveal a
Collaborator--Assistant spectrum (Cram\'er's V~$= 0.50$):
agents initiate work, yet terminal merge authority remains human.
These findings represent a snapshot of AI-assisted development
during its early production maturity (2025~Q2--Q3);
longitudinal replication will be necessary to track how these
patterns evolve.
Teams and researchers can use the taxonomy and five state machines
as a reusable framework for review allocation and tool selection.

The 54 automation-authorized merges (0.24\% of merged PRs) expose a measurement boundary between merge \emph{mechanism} and governance \emph{decision}; resolving it requires live platform API data (Section~\ref{sec:discussion}).

\begin{acks}
We acknowledge the support of the Natural Sciences and Engineering Research Council of Canada (NSERC), RGPIN-2021-03969.
\end{acks}

\bibliographystyle{ACM-Reference-Format}

\end{document}